\documentclass[traditabstract]{aa}
\usepackage[varg]{txfonts}
\usepackage{graphicx}
\usepackage{natbib}
\usepackage{amsmath,amssymb}
\usepackage{hyperref}
\usepackage{rotating, bm}

\newcommand{\mgk}{\ion{Mg}{ii}\,k}

\begin{document} 

\title{Solar spicule population properties inferred from IRIS Mg~II~k off-limb spectra using likelihood-free Bayesian inference}

\titlerunning{Statistical modeling of solar spicules}

\author{
Ji\v{r}\'{\i} \v{S}t\v{e}p\'an\inst{1}
\and
Stanislav Gun\'ar\inst{1}
\and
Akiko Tei\inst{2}\thanks{Research Fellow of Japan Society for the Promotion of Science}
\and
Petr Heinzel\inst{1,3}
}

\institute{
Astronomical Institute of the Academy of Sciences, 25165 Ond\v{r}ejov, Czech Republic.
\and
National Astronomical Observatory of Japan, 2-21-1 Osawa, Mitaka, Tokyo 181-8588, Japan
\and
Center of Excellence---Solar and Stellar Activity, University of Wroclaw, Kopernika 11, 51 622 Wroclaw, Poland
}

\authorrunning{\v{S}t\v{e}p\'an et al.}

\date{Received XXXX; accepted XXXX}

\abstract{
Solar spicules are ubiquitous and strongly overlap in off-limb observations, which complicates the interpretation of individual events. We aim to infer population-level properties of spicules from IRIS Mg~II~k off-limb spectra using a forward model that explicitly accounts for line-of-sight superposition and instrumental effects. We constructed a stochastic 3D spicule-population model and synthesized Mg~II~k spectra with radiative transfer, including instrumental degradation and photon noise. We then constrained the hyperparameters that describe the underlying spicule population with likelihood-free Bayesian inference. This approach combines approximate Bayesian computation and an adaptive population Monte Carlo sampler, comparing observed and simulated datasets through principal-component summary statistics at five heights above the limb. The inferred posterior constrains effective population-level properties of the spicules that dominate the observed Mg~II~k variability and height dependence. These constraints provide a route to connect off-limb spectroscopy with forthcoming spectropolarimetric diagnostics of the chromosphere--corona interface.
}

\keywords{
radiative transfer ---
Sun: chromosphere ---
Sun: corona ---
Sun: UV radiation
}

\maketitle

\section{Introduction
\label{sec:intro}}

\begin{figure*}
\centering
\includegraphics[width=1\hsize]{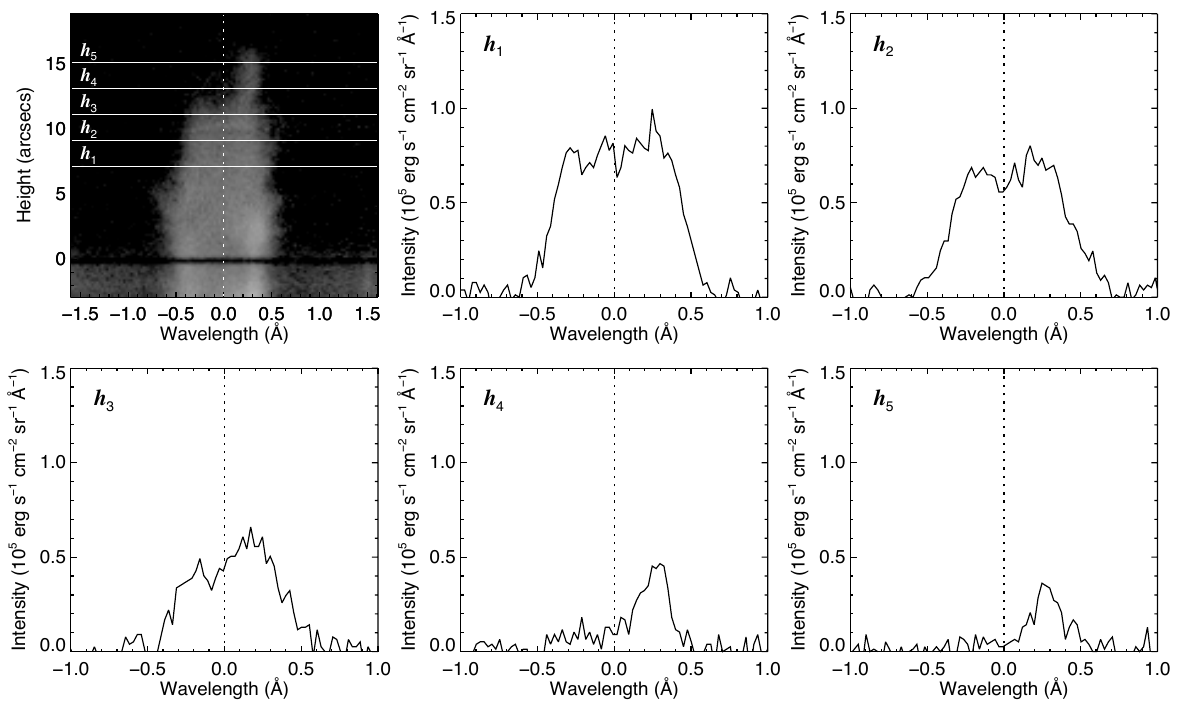}
\caption{
Example of the \mgk\ line IRIS observation on 2016/02/21 at 12:51:04.770. Top left panel: Line spectra observed along the slit perpendicular to the solar limb. The five horizontal white lines indicate the heights above the limb chosen for our analysis. The remaining panels, indicated by $h_1$ to $h_5$, show the spectra at these respective heights.
The selected heights are $h_1=4.6^{\prime\prime}$, $h_2=6.6^{\prime\prime}$, $h_3=8.6^{\prime\prime}$, $h_4=10.7^{\prime\prime}$, and $h_5=12.7^{\prime\prime}$ above the photospheric level.
As height increases, the line profile results from fewer individual Doppler-shifted spicule contributions.
}
\label{fig:obs}
\end{figure*}

Solar spicules are thin, elongated, short-lived, and highly dynamic structures present in large numbers all around the Sun at any given time. They can be observed anywhere on the solar surface and in emission all around the solar limb. Spicules are believed to be primarily chromospheric temperature phenomena, even though they extend into the low corona and are intimately connected to the chromosphere--corona interface. Even though spicules have been known to exist on the Sun for over a century, their properties are still not well understood. Spicules owe their enigmatic nature to their small dimensions, which remain near the resolution limit of the current instruments. Moreover, observations of spicules are further complicated by their highly dynamic behavior. 

Due to their ubiquitous presence and varying inclinations, there is a high probability of observing many different spicules along a given line of sight (LOS) at lower altitudes above the solar limb. Spicules there form a dense forest. Observations and analyses of individual spicules are thus typically constrained only to those spicules that stand out from the background. However, analyses of individual examples come with the inherent risk that the analyzed spicules may not fully represent the properties of solar spicules in general.

In the following paragraphs, we present a short overview of certain spicule properties subject of this study. A comprehensive review of spicules can be found in \citet{2012SSRv..169..181T}. The observations of spicules reveal a highly dynamic nature of this phenomenon. We can clearly see fast motions along the length of spicules (both toward and away from the solar surface) as well as considerable motions sideways and toward or away from the observer. The velocities along the length of spicules, typically measured in the plane of the sky (POS) from high-cadence imaging observations, generally range from 15 to 40 km s$^{-1}$ but may reach up to 100 km s$^{-1}$ \citep{2007Sci...318.1574D,2012ApJ...759...18P,2012ApJ...750...16Z}. For example, \citet{2014ApJ...792L..15P} derived upward velocities of around 80 km s$^{-1}$ and downward velocities reaching 140 km s$^{-1}$. Velocities perpendicular to the spicule axis measured in the POS are typically between 5 and 30 km s$^{-1}$ \citep{2007Sci...318.1574D,2012ApJ...759...18P,2011ApJ...736L..24O}. The measured LOS velocities toward and away from the observer range from $\pm 10$ to $\pm 30$ km s$^{-1}$ \citep{2009SoPh..260...59P,2014ApJ...795L..23S}. Observational analyses mostly focus on individual, well-resolved spicules that typically reach higher altitudes. Thus, a selection effect that could favor spicules with a certain set of parameters may play a role \citep{2012ApJ...759...18P}. To overcome this observational bias, \citet{2020ApJ...888...42T} analyzed a large dataset of \ion{Mg}{ii}\,h and k spectral observations of spicules obtained by the Interface Region Imaging Spectrograph \citep[IRIS;][]{2014SoPh..289.2733D}. These authors used 1D NLTE (i.e., departures from local thermodynamic equilibrium) models in multi-slab configuration to study overlapping spicules and find that configurations with LOS velocities ranging from $-25$ to 25~km\,s$^{-1}$ can reproduce the observed spectra. Non-LTE modeling of IRIS observations in multiple spectral lines (\ion{Mg}{ii}\,h and k, \ion{C}{ii}, and \ion{Si}{iv}) was also performed by \citet{2018SoPh..293...20A} using a 1D single-slab geometry.

The picture of spicule dynamics is further complicated by the wide range of inclinations that they exhibit. Because the inclination component leaning toward or away from the observer is unknown, all measured velocities represent a projection of the actual velocity vector. The spicule inclination from the local vertical is closely connected to the orientation of the local magnetic field. The typically measured values of inclination are around 25$^\circ$ from the vertical \citep{2009SoPh..260...59P,2012ApJ...759...18P}. However, it should be noted that these measurements derive only the projection of the actual tilt of a spicule into the POS (a plane normal to the LOS). This naturally leads to an underestimation of the true inclination. Another source of a bias toward smaller-than-actual values of spicule inclination is the selection effect that favors observations of spicules that protrude to higher altitudes. In fact, a vast majority of spicules with larger inclinations would not have sufficient lengths to rise above the canopy of the spicular forest and thus go mostly undetected. For example, an analysis of spectro-polarimetric observations by \citet{2005ApJ...619L.191T} derived a magnetic field inclination in spicules of around 35$^\circ$ from the local vertical.

The number of spicules typically occurring on the Sun at any given time is estimated to be above 10$^{6}$, with some studies suggesting higher numbers of up to $2 \times 10^{7}$ \citep{2012SSRv..169..181T}. However, these estimates strongly depend on the quality of observations and the altitude at which the spicules are counted.

Measurements of the dimensions of spicules also strongly depends on the quality of the observations as well as instrument resolution. Spicule length is generally measured as the distance from the photosphere to the point at which a spicule becomes invisible in the spectral line used for observations. Such measurements naturally represent a projection of the actual length of spicules onto the POS. For example, high-resolution space-borne observations by the Solar Optical Telescope \citep[SOT;][]{2008SoPh..249..167T} on board the Hinode satellite \citep{2007SoPh..243....3K} in the Ca II H line show lengths ranging from a few hundred to 10,000 km, with the majority below 5,000 km \citep{2007PASJ...59S.655D}. The spicule widths measured in these observations range from 200 km to 1000 km \citep{2011NewA...16..296T}.

While the presence of a large number of overlapping spicules along a given LOS crossing the spicular forest complicates the analysis of individual structures, it is advantageous for statistical analyses. The essential assumption of our approach is that the physical parameters of spicules are sampled from an unknown probability distribution characteristic of the observed conditions, i.e., a given magnetic environment and a given phase of the solar cycle. Our goal in this paper is to take a first step toward inferring this distribution.

The paper is organized as follows. In Sect.~2 we describe the IRIS observations and the preparation of the off-limb \mgk\ spectra at the selected heights. Section~3 introduces the spicule-population forward model and summarizes the radiative-transfer assumptions. In Sect.~4 we present the likelihood-free inference approach, including the choice of summary statistics, discrepancy, and the adaptive population Monte Carlo (APMC) algorithm. The inferred posteriors are discussed in Sect.~5. Finally, Sect.~6 summarizes the main conclusions, limitations, and prospects for future improvements.

\section{The IRIS observations of the coronal-hole spicules
\label{sec:formul}}

This section provides an overview of the observed data used later for inference, and a qualitative description of how the synthetic spectra are formed. All technical details, algorithms, and parameterizations are described in Sects.~\ref{sec:model} and \ref{sec:inference}.

We used a broad dataset of high-cadence spectroscopic observations of off-limb spicules obtained in the \mgk\ line by IRIS. The data consist of over 37,000 individual \mgk\ profiles observed in a polar coronal hole on Feb 21, 2016. The spectra were taken between 12:29 and 13:29 UT with a very short exposure time of 5.4 s. This freezes the movements of the highly dynamic spicules, allowing us to study the properties of spicules unaffected by averaging over temporal variations. The data were obtained in the sit-and-stare mode at the southern polar region. The spatial pixel size is 0.17 arcsec, and the spectral pixel size is 25.6 m\AA. More details about the used dataset can be found in \citet{2020ApJ...888...42T}. 

The studied spectra exhibit a significant variation with altitude. The \mgk\ profiles obtained near the solar limb are broad and generally double-peaked; profiles at midlatitudes are broad and often flat-topped; and profiles at higher altitudes are typically narrower and often single-peaked (for more details, see Fig.~2 of \citealt{2020ApJ...888...42T}). This complex behavior, which can be also seen in other \mgk\ data of spicules \citep[e.g., in][]{2025ApJ...983..108T}, is a consequence of the overlapping forest of highly dynamic spicules. Individual spicules can be observed only when protruding above the canopy, while the spectra obtained from within the forest are formed by multiple overlapping spicules exhibiting large (and essentially random) variations in LOS velocities. Such an environment is challenging to study through a comparison of observed spectra with synthetic spectra produced by NLTE modeling using individual realizations of modeled structures. Nonetheless, it represents an advantage for studies using sophisticated statistical techniques.  

Following the selection shown in Fig.~\ref{fig:obs}, we extracted the \mgk\ spectra at five fixed heights above the photospheric limb, labeled $h_1$ to $h_5$. Our statistical analysis was performed on the ensemble of 250 profiles at each height, and the constraints reported below combine information from all five height samples. At the lower heights in particular, each profile corresponds to radiation formed along a LOS through a dense spicule forest, so that the observed signal is due to the superposition of many spicules. Depending on height and local crowding, a single LOS can intersect hundreds and, in the densest part of the forest, even a few thousand spicules, making it impractical to interpret the spectra in terms of isolated individual structures.

\section{Forward model
\label{sec:model}}

\begin{figure}
\centering
\includegraphics[width=1\hsize]{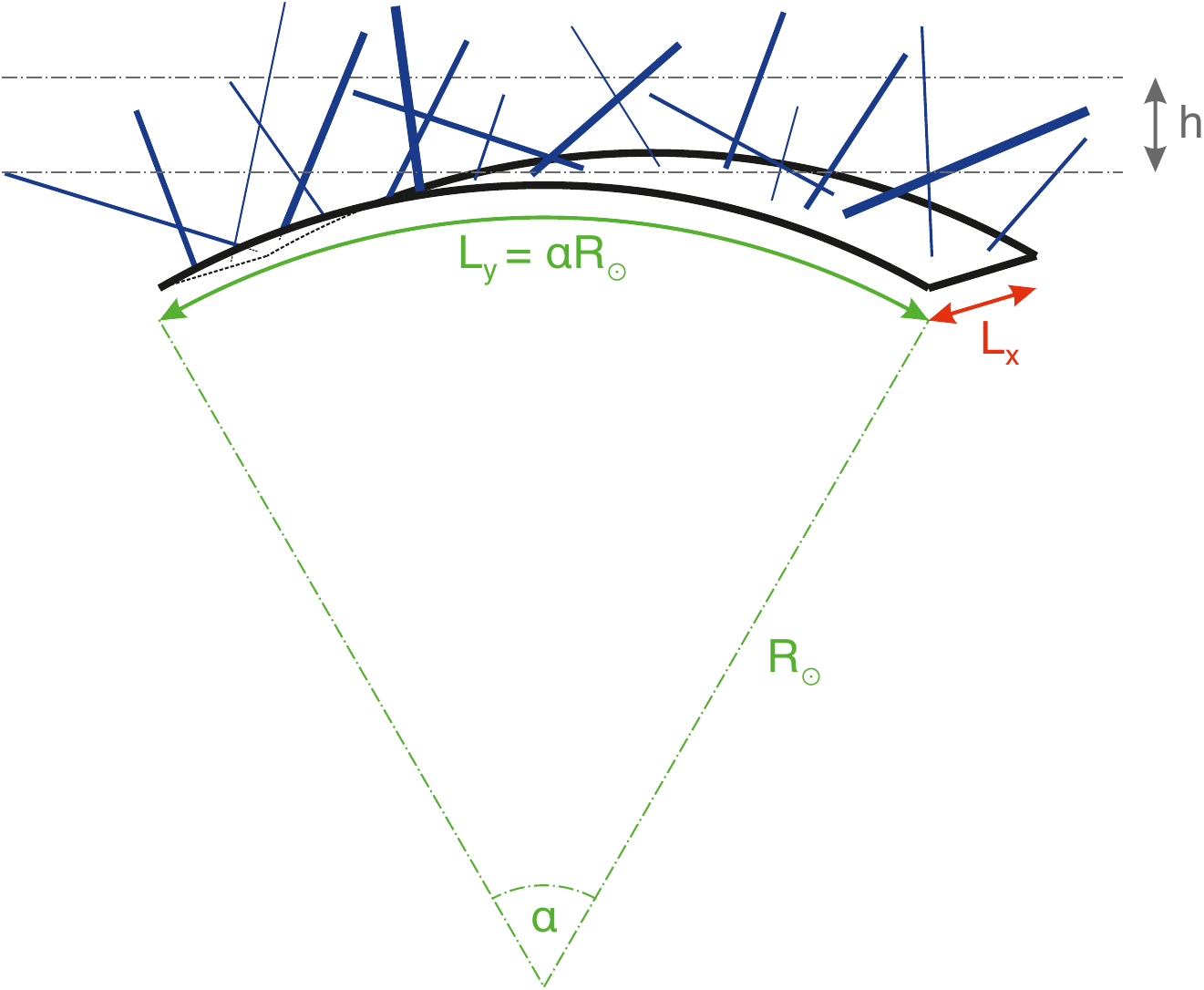}
\caption{
Schematic drawing demonstrating a random realization of the spicule forest (blue lines) and the observing geometry.
The spectrograph slit is parallel to the arrow defining $h$.
See text for details.
}
\label{fig:forest}
\end{figure}

\begin{figure}
\centering
\includegraphics[width=0.8\hsize]{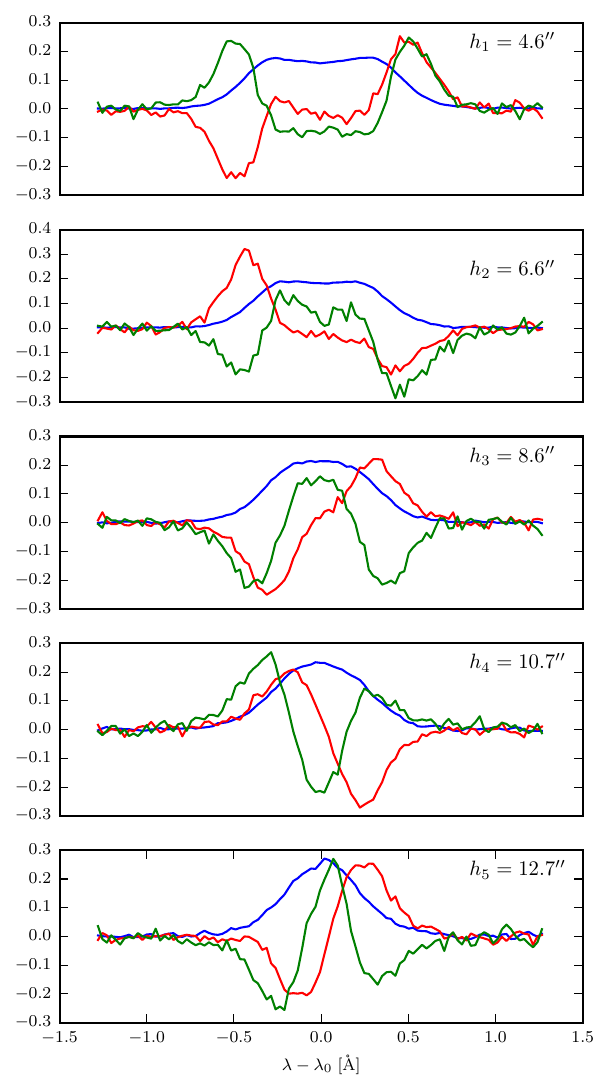}
\caption{
First three PCA eigenprofiles $e_{i,k}(\lambda)$ of the observed \mgk\ line profiles at each analyzed height ($h_i$) above the limb. The colors represent the component index: $k=1$ (blue), $k=2$ (red), and $k=3$ (green). These height-dependent bases are used to project individual profiles and obtain the coefficients ($a_{i,k}$), whose ensemble moments define the summary vector $S(\mathcal{D})$ used in the ABC discrepancy.
}
\label{fig:eigen}
\end{figure}

\subsection{Geometric model of the spicule population
\label{ssec:pop_geom}}

We modeled a 3D ensemble of spicules as straight circular cylinders, whose bases are distributed on a narrow stripe of the solar surface extending around the limb, i.e., including regions slightly in front of and behind the POS. The stripe is parameterized by $(x,y)\in[-L_x/2,L_x/2]\times[-L_y/2,L_y/2]$, where $x$ is the tangential coordinate parallel to the POS and $y$ is the linear arc-length coordinate measured around the limb toward or away from the POS (see the schematic drawing in Fig.~\ref{fig:forest}). The total span in this second direction is $L_y=\alpha R_\sun$, with $R_\sun$ the solar radius and $\alpha$ the angular separation between the stripe ends. The spectrograph slit was centered at $x=0$, and the coordinate along the slit is the height $h$ above the limb. The width $L_x$ was chosen so that spicules with lengths within the prior support can intersect the slit at the sampled heights, and $L_y$ was chosen to be wide enough to include bases from both near-front and near-back hemispheres that can contribute along the LOS.

The number of spicules placed on the stripe was set by the model’s surface number density $A$ (in $\mathrm{Mm}^{-2}$; defined below). For each realization, we placed $N(A)=\lceil A L_x L_y \rceil$ spicules, and we drew base locations randomly over $(x,y)$ on the stripe. The properties of individual spicules (i.e., spicule orientation ($\mu,\phi$), effective length $\ell$, effective diameter $D$, etc.) were independently sampled from population distributions governed by hyperparameters of the model. The exact forms of these distributions and their priors are specified in Sect.~\ref{sec:inference}. The azimuthal orientations were uniformly drawn around the local vertical. A schematic random realization is shown in Fig.~\ref{fig:forest}.

For each height window $h_i$, we identified the segment of every cylinder that intersects the LOS at a given height $h_i$ and computed its radiative contribution. Along a given LOS, contributions from different spicules were combined, as described in Sect.~\ref{ssec:rt}. In practice, only a small fraction of simulated spicules intersects the LOS in any given height window.

\subsection{Physical assumptions and radiative transfer
\label{ssec:rt}}

To construct a statistical model of spicules, we adopted several physical approximations that make the problem tractable in practice. Below we describe our assumptions regarding spicule properties and the treatment of radiative transfer (RT). Specific parameterizations and priors are provided later in the text (see Sect.~\ref{ssec:param_priors}).

We considered a two-level model atom for the synthesis of \mgk, the ${}^2\mathrm{P}_{3/2}$–${}^2\mathrm{S}_{1/2}$ transition, and neglect collisions. We assumed that the individual spicules are sufficiently optically thin, so self-consistent RT within a spicule can be neglected. This assumption was validated a posteriori by checking that the inferred parameters imply line-center optical thicknesses below unity. If individual spicules were optically thick, the modeling framework would need a revision.

The line source function, $S$, was determined by the mean incident radiation $\overline{J}$ from the underlying chromosphere. For a given height and plasma velocity vector, $\overline{J}$ was computed by integrating the Doppler-shifted specific intensity of the semiempirical FAL-C model \citep{1993ApJ...406..319F} over the solar disk and the line absorption profile. To accelerate the solution, we used a simple neural network surrogate that predicts $\overline{J}$ from three inputs: height above the chromosphere, the plasma velocity vector, and the local Doppler width.

Because spicules are spatially separated, we neglected mutual radiative interaction among different spicules, an approximation which has been adopted in off-limb \mgk\ studies in the past \citep[e.g.,][]{2020ApJ...888...42T}. As in these studies, we integrated the RT equation through the forest of spicules (see below). This approximation is delicate and can break down if the surface number density increases, in which case inter-spicule coupling would require a more complex NLTE treatment. We return to this issue in Sects.~\ref{sec:results} and \ref{sec:concl}.

Along each LOS we solved the RT by accounting for all intersected spicules. Inside a single spicule the monochromatic opacity is modeled as
\begin{equation}
\chi(\lambda,r)=\chi_{\rm L}(r)\exp\left(-\frac{(\lambda-\lambda_{\rm L})^2}{\Delta\lambda_{\rm D}^2}\right),
\end{equation}
where $r$ is the radial distance from the spicule axis, $\Delta\lambda_{\rm D}$ is the line Doppler width, and $\lambda_{\rm L}=\lambda_0\,(1+v_{\parallel}/c)$ is the Doppler-shifted line center wavelength given the LOS velocity $v_{\parallel}$ (the projection of the local plasma velocity onto the LOS). The radial dependence of the opacity scale was taken to be Gaussian,
\begin{equation}
\chi_{\rm L}(r)=\chi_{\rm L}\,\exp \left(-\frac{r^2}{2R^2}\right),
\end{equation}
with $R$ being the Gaussian scale radius (or effective radius) and $\chi_{\rm L}$ the line opacity expressed in inverse megameters. We denote by $D$ the effective diameter or spatial full width at half maximum (FWHM) of this cross-section, where $D = 2\sqrt{2\ln 2}\,R$. The LOS optical depth at wavelength $\lambda$ is then
\begin{equation}
\tau(\lambda)=\int_{\mathcal{C}} \chi(\lambda,r)\,ds
= \exp\left(-\frac{(\lambda-\lambda_{\rm L})^2}{\Delta\lambda_{\rm D}^2}\right)\,
\int_{\mathcal{C}} \chi_{\rm L}(r)\,{\rm d}s,
\end{equation}
where $\mathcal{C}$ is the LOS chord through the spicule. For a perpendicular LOS through the spicule center, the line-center optical thickness evaluates to
\begin{equation}
\tau_0=\int_{-\infty}^{\infty} \chi_{\rm L} \exp\left(-\frac{r^2}{2R^2}\right)\,{\rm d}r
= \chi_{\rm L}\sqrt{2\pi}\,R \approx \chi_{\rm L}\,D.
\label{eq:tau0}
\end{equation}
When the spicule axis is inclined relative to the LOS, the chord length increases, and the LOS optical thickness grows accordingly.

Let us assume that a given LOS ray intersects a
sequence of spicules labeled $j=1,\ldots,n$. If the intensity
incident on spicule $j$ along the LOS is $I_{j-1}(\lambda)$, the line source function is $S_j$, and the optical thickness along $\mathcal{C}$ is $\tau_j(\lambda)$. Thus, the emergent intensity from this spicule is
\begin{equation}
I_j(\lambda)=
I_{j-1}(\lambda)e^{-\tau_j(\lambda)}
+
\left(1-e^{-\tau_j(\lambda)}\right)S_j\,.
\end{equation}
Applying this relation successively from $j=1$ and assuming $I_0(\lambda)=0$, we obtain the following expression for the intensity emerging from the $k$-th spicule:
\begin{equation}
I_k(\lambda)
=
\sum_{j=1}^{k}
S_j
\left(1-e^{-\tau_j(\lambda)}\right)
e^{-\sum_{i=j+1}^{k}\tau_i(\lambda)}\,.
\end{equation}
This expression can be used to compute the total emergent intensity from the full forest of spicules by setting $k=n$.

The effective diameter $D$ and effective length $\ell$ considered here are effective quantities tied to the formation of \mgk. They reflect the parts of a spicule that contribute most strongly to the \mgk\ contribution function under our radiative–transfer assumptions (through $\chi_{\rm L}$, $\Delta\lambda_{\rm D}$, and LOS integration), as seen after convolution with the instrumental point spread function (PSF). Diagnostics formed at different temperatures and heights and with different spectral lines would generally yield different apparent or effective diameters and lengths. Thus, $D$ and $\ell$ should be interpreted as line-dependent effective descriptors.

\subsection{Velocities and kinematics
\label{ssec:kinematics}}

For inference we kept the kinematics minimal. We modeled only axial motions and assumed a constant axial speed $v$ along each spicule. No along–spicule gradients were included (neither in $v$ nor in other quantities). We parameterized $v$ as a non–negative speed. The sign of $v_{\parallel}$ was then set by the random orientation of the spicule axis.\footnote{With azimuths uniformly drawn around the local vertical, reversing the flow direction corresponds to an equivalent ensemble under a change of azimuth. Thus, it suffices to parameterize the speed magnitude.}

Non–axial motions (swaying or transverse and torsional) were not modeled explicitly, although they are commonly observed in spicules \citep[e.g.,][]{2012ApJ...752L..12D}. Their unresolved contribution was absorbed into the Doppler width $\Delta\lambda_{\rm D}$ used in the absorption profile, so that $\Delta\lambda_{\rm D}$ represents the combined effect of thermal and unresolved dynamical broadening under our observing conditions.

\subsection{Instrumental degradation of the spectra
\label{ssec:inst_noise}}

To compare simulations with the IRIS data, we degraded the synthetic spectra to match the instrumental resolution and noise properties of the observations. First, high-resolution model profiles were convolved with the IRIS spectral line-spread function, which we approximated by a Gaussian with ${\rm FWHM}_{\lambda}=50.54\,\text{m\AA}$ \citep{2014SoPh..289.2733D}. After convolution, spectra were resampled to the observed wavelength grid.

Next, we added Poisson noise at a level estimated from the data (using line wings and continuum windows). Practically, model intensities are scaled to effective photon counts per spectral bin, Poisson deviates are drawn, and the results are converted back to radiometric units, preserving the observed signal-to-noise characteristics.

\section{Bayesian inference and approximate Bayesian computation
\label{sec:inference}}

\subsection{Hierarchical Bayesian modeling overview
\label{ssec:bayes_overview}}

Let $\mathcal{D}$ denote the observed dataset, that is, the collection of \mgk\ spectra extracted at the five heights $\{h_i\}$, described in Sect.~\ref{sec:formul}. Our forward model (Sect.~\ref{sec:model}) is a stochastic simulator. For a given set of population-level hyperparameters $\vec{\theta}$, it draws a particular realization of the spicule population, described by a set of latent (nuisance) spicule-level variables $\vec{\xi}$ (including the random locations of spicules and their individual properties). It then produces synthetic spectra through the radiative-transfer calculation followed by instrumental degradation and noise.

In a hierarchical Bayesian formulation \citep[see, e.g.,][]{2014A&A...572A..98A}, the joint posterior for the latent variables and hyperparameters is
\begin{equation}\label{eq:hier_post}
p(\vec{\xi},\vec{\theta} | \mathcal{D})\propto p(\mathcal{D} | \vec{\xi},\vec{\theta})p(\vec{\xi} | \vec{\theta})\pi(\vec{\theta}),
\end{equation}
where $\pi(\vec{\theta})$ is the hyperprior and $p(\vec{\xi}| \vec{\theta})$ encodes the population model. We are primarily interested in the hyperparameters, so we marginalized over the latent spicule realization,
\begin{equation}\label{eq:marg_post}
p(\vec{\theta}| \mathcal{D})=\int p(\vec{\xi},\vec{\theta}| \mathcal{D})\,{\rm d}\vec{\xi}\propto \pi(\vec{\theta})p(\mathcal{D} | \vec{\theta}),
\end{equation}
with the marginal likelihood
\begin{equation}\label{eq:marg_like}
p(\mathcal{D}| \vec{\theta})=\int p(\mathcal{D}| \vec{\xi},\vec{\theta})p(\vec{\xi}| \vec{\theta})\,{\rm d}\vec{\xi}.
\end{equation}
In our case, $p(\mathcal{D} | \vec{\xi},\vec{\theta})$ is not available in closed form, and Eq.~\ref{eq:marg_like} is prohibitively high-dimensional, which motivates likelihood-free inference via approximate Bayesian computation (ABC).

The APMC sampler does not explore the joint $(\vec{\xi},\vec{\theta})$ space. Its particles contain only the hyperparameters $\vec{\theta}$, while a new realization $\vec{\xi}\sim p(\vec{\xi}\mid\vec{\theta})$ is generated internally for each forward simulation. The latent variables were therefore marginalized over by simulation rather than included in the sampled parameter space.

\subsection{Parameterization and population priors
\label{ssec:param_priors}}
The simulator is parameterized hierarchically. The hyperparameters $\vec{\theta}$ control the population-level distributions from which the individual spicule properties are drawn, while the spicule-level parameters $\vec{\xi}$ describe one particular realization of the population. The hyperparameters inferred by ABC are summarized in Table~\ref{tab:hyper}, and the conditional priors used to draw the spicule-level parameters are listed in Table~\ref{tab:prior}.

\begin{table*}
\caption{Hyperparameters and their hyperpriors.}
\label{tab:hyper}
\centering
\begin{tabular}{c l c}
\hline\hline
Hyperparameter & \multicolumn{1}{c}{Description} & Hyperprior \\
\hline
$A$ & surface number density of spicules & $\mathcal{J}(A)$ \\
$\alpha_\mu, \beta_\mu$ & distribution of inclinations & $\propto (\alpha_\mu+\beta_\mu)^{-5/2}$ \\
$\alpha_v, \beta_v$ & distribution of velocities & $\propto (\alpha_v+\beta_v)^{-5/2}$ \\
$\Delta\lambda_{\rm D}$ & absorption profile width & $\mathcal{U}(0\,\mathrm{\AA},1\,\mathrm{\AA})$ \\
$\chi_{\rm L}$ & line opacity & $\mathcal{J}(\chi_{\rm L})$ \\
$\sigma_\ell$ & mode of the effective length  & $\mathcal{J}(\sigma_\ell)$ \\
$\sigma_R$ & mode of the effective radius & $\mathcal{J}(\sigma_R)$ \\
\hline
\end{tabular}
\tablefoot{
The last column lists the hyperprior densities used to draw proposed hyperparameter values: $\mathcal{U}(a,b)$ denotes a uniform distribution on $[a,b]$, $\mathcal{J}(x)$ denotes a Jeffreys (log-uniform) prior for a positive scale parameter ($p(x)\propto x^{-1}$), and the entries $\propto (\alpha+\beta)^{-5/2}$ specify the joint hyperprior for the beta-distribution shape parameters with $\alpha,\beta>0$ (given up to a normalization constant).
}
\end{table*}

\begin{table*}
\caption{Prior distributions for the spicule-level parameters ($\vec{\xi}$).}
\label{tab:prior}
\centering
\begin{tabular}{c l c}
\hline\hline
Symbol & \multicolumn{1}{c}{Quantity} & Prior \\
\hline
$x,y$ & spicule location ${\rm [Mm,Mm]}$ & $\mathcal{U}(-L_x/2,L_x/2), \mathcal{U}(-L_y/2,L_y/2)$ \\
$\phi$ & spicule azimuth {\rm [rad]} & $\mathcal{U}(0,2\pi)$ \\
$\mu$ & cosine of the spicule inclination & ${\rm Beta}(\alpha_\mu,\beta_\mu)$ \\
$v$ & longitudinal velocity & $v_{\rm max}{\rm Beta}(\alpha_v,\beta_v)$ \\
$\lambda_w$ & absorption profile width $[\AA]$ & $\delta(\lambda_w-\Delta\lambda_{\rm D})$ \\
$\chi$ & line opacity $[{\rm Mm^{-1}}]$ & $\delta(\chi-\chi_{\rm L})$ \\
$\ell$ & effective length $[{\rm Mm}]$ & $\mathcal{R}(\sigma_\ell)$ \\
$R$ & effective radius $[{\rm Mm}]$ & $\mathcal{R}(\sigma_R)$ \\
\hline
\end{tabular}
\tablefoot{Some parameters are sampled conditionally based on the hyperparameters in Table~\ref{tab:hyper}; others are tied deterministically to their hyperparameters. In the notation, ${\rm Beta}(\alpha,\beta)$ denotes a beta prior, $\mathcal{R}(\cdot;\sigma)$ a Rayleigh prior, and $\delta(\cdot)$ a Dirac delta.}
\end{table*}

The surface number density ($A$) sets the expected number of spicules per unit area in the POS domain used by the simulator (see Fig.~\ref{fig:forest}). The pairs $(\alpha_\mu,\beta_\mu)$ and $(\alpha_v,\beta_v)$ are the shape parameters of the beta distributions adopted for the inclination and velocity distributions, respectively. For the inclination, we followed the hierarchical parameterization discussed by \citet{2014A&A...572A..98A}. For the velocities, we took advantage of the flexibility of the beta distribution to accommodate a range of distribution shapes. The remaining hyperparameters specify the absorption-profile width and the line opacity as well as the scale parameters of the Rayleigh distributions used for the apparent spicule lengths and radii. The corresponding hyperpriors are listed in the last column of Table~\ref{tab:hyper}. We assumed that the full hyperprior $\pi(\vec\theta)$ factorizes over these parameter blocks, that is, all hyperparameters are a priori independent except for the jointly specified pairs $(\alpha_\mu,\beta_\mu)$ and $(\alpha_v,\beta_v)$. We chose broad ranges to keep them weakly informative. Except for the Doppler width, for which we assumed a uniform prior, we mostly adopted Jeffreys (log-uniform) priors for scale parameters. As we show later, the inferred posteriors differ clearly from the priors, indicating that the observations provide meaningful constraints on the model hyperparameters.

Conditioned on the hyperparameter vector $\vec{\theta}$, the simulator draws the spicule-level parameters $\vec{\xi}$ independently for each spicule, following the conditional priors summarized in Table~2. Concretely, the spicule base location is sampled uniformly over the POS stripe,
\begin{equation}
x\sim \mathcal{U}(-L_x/2,L_x/2),\qquad y\sim \mathcal{U}(-L_y/2,L_y/2),
\end{equation}
and the azimuth is isotropic,
\begin{equation}
\phi\sim \mathcal{U}(0,2\pi).
\end{equation}
The inclination is parameterized by $\mu=\cos\vartheta$ and sampled as
\begin{equation}
\mu\sim \mathrm{Beta}(\alpha_\mu,\beta_\mu),
\end{equation}
while the axial speed is nonnegative and bounded by construction,
\begin{equation}
\frac{v}{v_{\max}} \sim \mathrm{Beta}(\alpha_v,\beta_v),
\end{equation}
with the sign of the LOS component set by the random orientation of the spicule axis. In practical calculations we set $v_{\max}=200\,\mathrm{km\,s^{-1}}$.
For the geometric sizes, we modeled the effective length $\ell$ and the Gaussian cross-sectional scale radius $R$ as
\begin{equation}
\ell\sim \mathcal{R}(\sigma_\ell),\qquad R\sim \mathcal{R}(\sigma_R),
\end{equation}
i.e., Rayleigh distributions with scale parameters $\sigma_\ell$ and $\sigma_R$.
Finally, the absorption-profile Doppler width and line opacity were taken to be constant across spicules within one realization,
\begin{equation}
\lambda_w\sim \delta(\lambda_w-\Delta\lambda_{\rm D}),\qquad \chi\sim \delta(\chi-\chi_{\rm L}),
\end{equation}
so that $\Delta\lambda_{\rm D}$ and $\chi_{\rm L}$ enter the model deterministically. These Dirac-delta choices reflect the simplifying assumption that the thermodynamic properties controlling the line width and opacity are uniform across spicules within a realization. Although idealized, this keeps the model simple.

For the apparent length and radius, we adopted Rayleigh distributions. This choice is primarily pragmatic: it enforces strict positivity, introduces only one scale parameter per quantity, and yields a right-skewed but light-tailed family that can represent broad variability without producing unrealistically frequent extreme outliers.

\subsection{Approximate Bayesian computation
\label{ssec:abc}}

\begin{figure*}
\centering
\includegraphics[width=0.85\hsize]{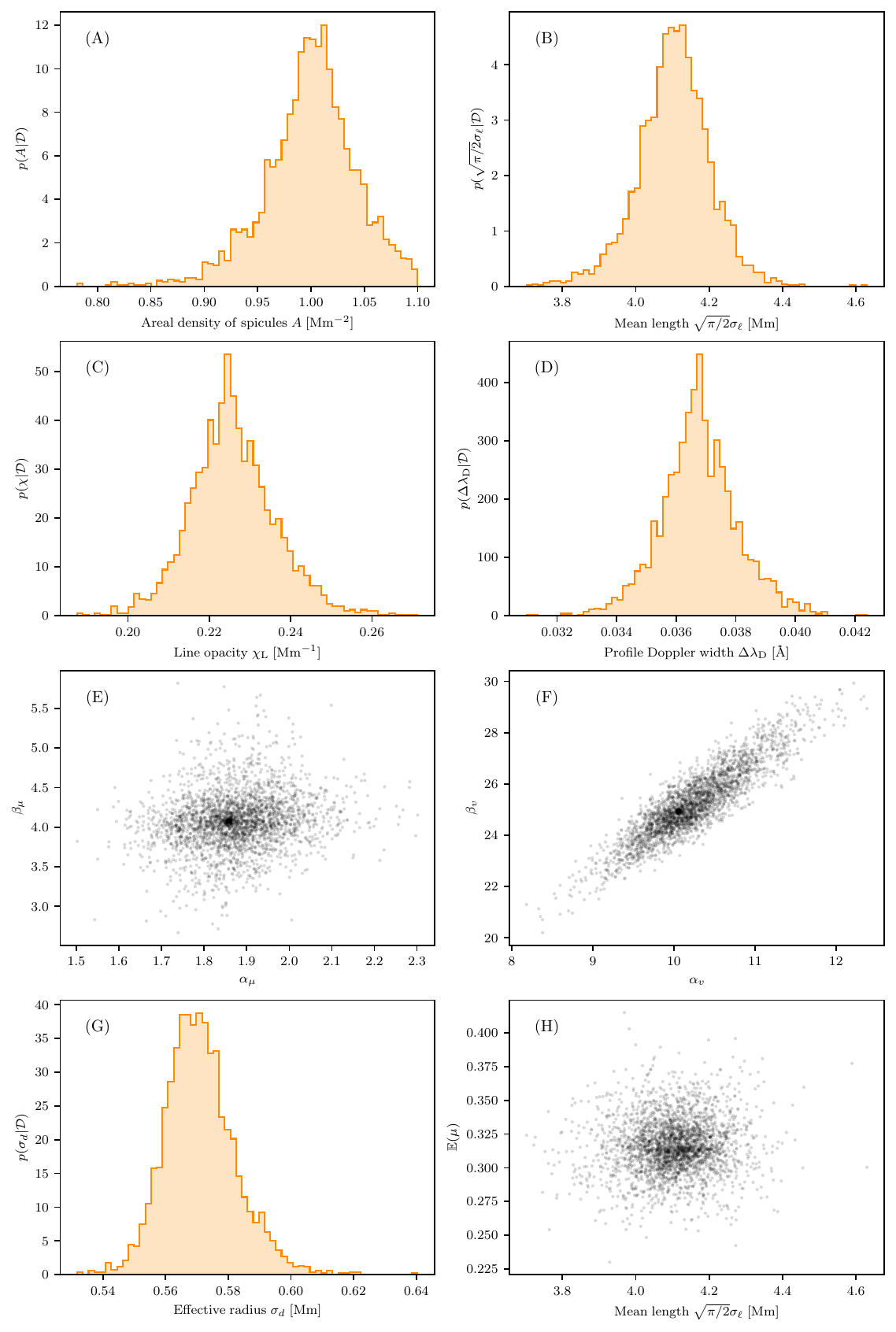}
\caption{
Marginal posteriors of hyperparameters (panels A--G).
Panel H: Distribution of the mean spicule length and the mean inclination of the spicules $\mathbb{E}(\mu)$. See text for details.
}
\label{fig:posthyper}
\end{figure*}

Approximate Bayesian computation replaces the intractable likelihood by comparing observed and simulated datasets \citep[e.g.,][]{AkeretEtAl2015ABC}. In its simplest rejection form, one draws $\vec{\theta}^\ast\sim\pi(\vec{\theta})$, simulates a dataset $\hat{\mathcal{D}}^\ast$ from the forward model at the same five heights as the observations, and accepts $\vec{\theta}^\ast$ if the discrepancy between $\mathcal{D}$ and $\hat{\mathcal{D}}^\ast$ is smaller than a tolerance $\varepsilon$. The accepted samples are then drawn from an approximate posterior $p_{\varepsilon}(\vec{\theta}\vert \mathcal{D})$. In the idealized setting in which the discrepancy compares the full data (or, equivalently, statistics that retain all information about $\vec{\theta}$), reducing the tolerance recovers the usual Bayesian posterior,
\begin{equation}
\lim_{\varepsilon\to 0} p_{\varepsilon}(\vec{\theta} \vert \mathcal{D})=p(\vec{\theta} \vert \mathcal{D}).
\end{equation}
Direct comparisons of individual profiles are not useful because the observed spectra at a given height form a large ensemble, and the correspondence between individual observed and simulated profiles is not defined. In practice, one typically compares lower-dimensional summary statistics $S(\mathcal{D})$ rather than the full dataset, which yields an approximation to the posterior that is conditioned on the chosen summaries. For fixed $S$, the zero-tolerance limit becomes
\begin{equation}
\lim_{\varepsilon\to 0} p_{\varepsilon}(\vec{\theta} \vert S(\mathcal{D}))=p(\vec{\theta} \vert S(\mathcal{D})),
\end{equation}
and the quality of the approximation to $p(\vec{\theta}\vert\mathcal{D})$ is controlled by how informative $S$ is for $\vec{\theta}$.

For each height $h_i$ we performed a principal component analysis (PCA) of the observed \mgk\ profiles at that height. Let $e_{i,k}(\lambda)$ denote the $k$th eigenprofile and let $a_{i,k}$ be the corresponding coefficient obtained by projecting an individual profile onto $e_{i,k}(\lambda)$. We retain the first $N_{\rm PCA}=5$ components and summarize the ensemble at height $h_i$ by the first two moments of these coefficients -- namely, their sample means and variances across all observed profiles at that height (see Fig.~\ref{fig:eigen}). Concatenating the summaries for the five heights yields the final summary vector $S(\mathcal{D})$. For each simulated dataset $\mathcal{\hat D}$, we compute $S(\mathcal{\hat D})$ in the same way, using the PCA bases $e_{i,k}(\lambda)$ derived from the observations.

The discrepancy between observed and simulated datasets is defined as the Euclidean distance between summary vectors,
\begin{equation}
\rho\big(S(\mathcal{D}),S(\mathcal{\hat D})\big)=\lVert S(\mathcal{D})-S(\mathcal{\hat D})\rVert_2 .
\end{equation}
We adopted this choice after testing alternatives because it captures the dominant profile-shape variability in a noise-robust way while keeping the ABC comparison low-dimensional.
We do not assume that these summaries are formally sufficient. The inferred posterior should therefore be understood as conditioned on the particular information retained by $S(\mathcal{D})$.
The choice of these summaries is pragmatic: they retain the dominant profile-shape variability and its height dependence while substantially reducing the dimensionality and sensitivity to noise. Posterior-predictive comparisons of the complete observed and simulated profile distributions would provide an additional validation of the information retained by these summaries and of the adequacy of the generative model as a whole.

Other choices are possible. For example, summary statistics can be learned from simulations by regressing the parameters on candidate features to construct low-dimensional, approximately sufficient summaries \citep[semiautomatic ABC;][]{FearnheadPrangle2012}, which can then be used with the same APMC sampler discussed in the following section.

\subsection{Adaptive population Monte Carlo sampler
\label{ssec:abcp_mc}}

Naive rejection sampling, described in the previous section, is inefficient in practice. Therefore, we approximate the posterior $p(\vec{\theta} | \mathcal{D})$ with the APMC ABC algorithm of \citet{2013Lenormand}. The method is sequential and likelihood-free: it constructs a series of populations of parameter samples that concentrate progressively toward regions of parameter space that produce simulated datasets closer to the observations, as measured by the discrepancy defined in Sect.~\ref{ssec:abc}. In this context, a particle is one Monte Carlo sample of the hyperparameter vector $\vec{\theta}$ together with its associated weight and the discrepancy $\rho$ obtained by running the forward model with that $\vec{\theta}$.

In our application, we used $N=2880$ particles and retained a fixed fraction $\alpha=0.7$ of the best particles at each iteration. Concretely, given a population $\{\vec{\theta}_t^{(i)}\}_{i=1}^N$ with associated discrepancies $\{\rho_t^{(i)}\}_{i=1}^N$, we set the tolerance $\varepsilon_t$ to the $\alpha$-quantile of the discrepancies and retained the $N_\alpha=\alpha N$ particles with the smallest distances. The remaining $N-N_\alpha$ particles were regenerated by perturbing the retained set:
\begin{enumerate}
\item We drew a seed particle $\vec{\theta}^\ast$ from the retained particles with probability proportional to its weight.
\item We proposed $\vec{\theta}'$ by perturbation with a multivariate normal kernel $K_t(\cdot\mid\cdot)$ whose covariance was adapted from the retained weighted sample \citep{2013Lenormand}.
\item We simulated a synthetic dataset $\mathcal{\hat D}'$ using the forward model (Sect.~\ref{sec:model}) and computed $\rho'=\rho\big(S(\mathcal{D}),S(\mathcal{\hat D}')\big)$.
\end{enumerate}
After generating the new particles, importance weights were updated to account for the perturbation kernel,
\begin{equation}
w_t^{(i)} = \frac{\pi\big(\vec{\theta}_t^{(i)}\big)}{\sum_{j=1}^{N_\alpha}\tilde w_{t-1}^{(j)}K_t\big(\vec{\theta}_t^{(i)} | \vec{\theta}_{t-1}^{(j)}\big)} .
\end{equation}
where $\tilde w_{t-1}^{(j)}$ are the normalized weights of the retained particles from the previous iteration.

The APMC algorithm is well suited to our problem because it (1) adapts the tolerance sequence directly from the simulated discrepancies (rather than relying on an ad hoc schedule), (2) increasingly focuses computational effort on regions of high posterior density as the tolerance decreases, and (3) provides an efficient sequential alternative to naive rejection ABC in moderate-to-high dimensions \citep{2013Lenormand}. We terminated the iterations once the fraction of newly generated particles dropped below 5\%, indicating that further reductions of the tolerance would require a disproportionate number of additional simulations.

\section{Results
\label{sec:results}}

We present posterior summaries for the hyperparameters of the spicule population and derived distributions for spicule quantities. Unless noted otherwise, results are shown as marginal posteriors in Fig.~\ref{fig:posthyper} and as induced distributions for physical parameters in Figs.~\ref{fig:mu}--\ref{fig:fwhm}). The uncertainties reflect the APMC posterior under the summary statistics defined in Sect.~\ref{ssec:abc} and the forward model of Sect.~\ref{sec:model}.

\subsection{Surface number density of spicules}

The published estimates of the surface number density of spicules, $A$, which are often reported as totals over the whole solar surface, differ substantially in the literature \citep[see, e.g.,][]{2012SSRv..169..181T}. For obvious reasons, it is difficult to obtain this number directly by counting spicules in a given spectral range. At both the limb and on the disk, spicules overlap, faint structures may be invisible at a given spatial resolution, and classification is imperfect. The surface number density derived from such observations should therefore be interpreted mainly as a lower limit to the true spicule count.

The marginal posterior for $A$ inferred in our calculations peaks around $1\,{\rm Mm^{-2}}$, with a very small variance (see panel~A of Fig.~\ref{fig:posthyper}). In terms of the total number of spicules on the solar surface at a given moment, this corresponds to about $6\times 10^{6}$ spicules, which is higher than the frequently quoted value of about $10^{6}$ but still below the maximum of $2\times 10^{7}$ reported by \citet{2012SSRv..169..181T}. As discussed below, this appears to be linked to the inclination distribution inferred here. If many spicules are more nearly horizontal, substantial overlap above the limb renders a large fraction unidentifiable, whereas the relatively small subset of more vertical spicules stands out as uniquely identifiable at the limb.

\subsection{Inclinations}

\begin{figure}
\centering
\includegraphics[width=0.8\hsize]{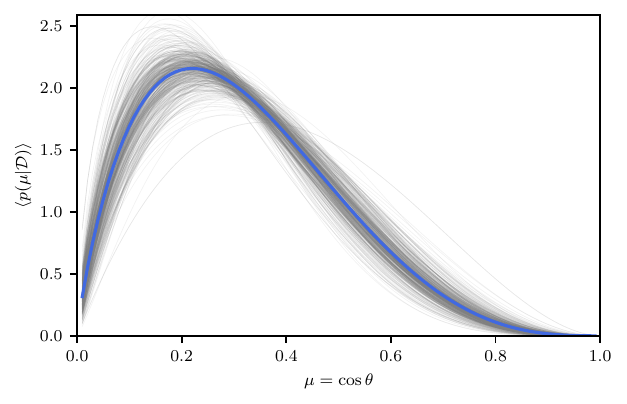}
\caption{
Posterior-induced distribution of the inclination cosine $\mu=\cos\vartheta$ with respect to the local solar vertical. The thick curve shows the posterior mean distribution, while the thin curves show distributions obtained from individual posterior draws of the hyperparameters. Here, $\mu=0$ and $\mu=1$ correspond to horizontal and vertical directions, respectively.
}
\label{fig:mu}
\end{figure}

\begin{figure}
\centering
\includegraphics[width=0.8\hsize]{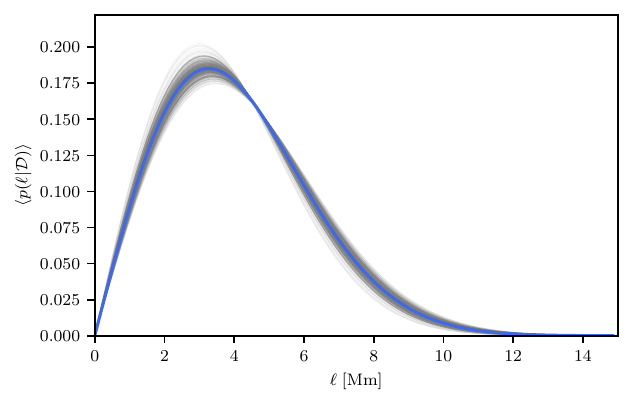}
\caption{
Distribution of the effective spicule length $\ell$ inferred from the \mgk\ data. As in Fig.~\ref{fig:mu}, the thick curve shows the posterior mean distribution, while the thin curves show distributions obtained from individual posterior draws of the hyperparameters.
}
\label{fig:ell}
\end{figure}

Following Table~\ref{tab:hyper}, we describe the spicule inclination with a population prior $\mathrm{Beta}(\alpha_\mu,\beta_\mu)$. The joint posterior for $(\alpha_\mu,\beta_\mu)$ (panel~E of Fig.~\ref{fig:posthyper}) concentrates on shapes that favor small $\mu$, i.e., a higher probability of more inclined (closer to horizontal) orientations. The induced distribution of $\mu$ is shown in Fig.~\ref{fig:mu}.

This result should be read in the observational context and in light of the fact that our observables come from higher heights above the solar surface (Fig.~\ref{fig:obs}). At these heights, near-vertical spicules stand out as separate features above the limb, whereas strongly inclined ones tend to overlap with neighbors, drop below the detection threshold, or be merged by the identification procedure. Consequently, our inference recovers the population that best reproduces the observed profiles at these heights rather than a complete census of all structures.
In this sense, the inclination distribution and related population parameters should be regarded as effective, model-dependent descriptors. In particular, the geometrically complex transition between the chromosphere and corona is represented only approximately in our simplified forward model, and we therefore do not interpret the inferred inclination distribution as a definitive measurement of the true solar spicule population.

Even though we refer to $A$ and the inclination distribution as effective, the inferred preference for small $\mu$ can still be given a physically consistent interpretation. In particular, if the low chromosphere contains a dense population in which spicules are on average more inclined than the subset that remains individually detectable at larger heights, then the near-limb spectra would be formed predominantly by LOS superposition of strongly overlapping, inclined structures. In that case, the population parameters inferred here should be understood as describing the subset of the forest that dominates the emergent \ion{Mg}{ii} signal in our height windows and under our implicit detection and background assumptions.

This interpretation is also compatible with the observed transition from broad, often double-peaked profiles close to the limb to narrower, often single-peaked profiles at larger heights \citep[see Fig.~2 of][and our Fig.~\ref{fig:obs}]{2020ApJ...888...42T}. In our model, the intrinsic absorption profile of an individual spicule is relatively narrow (set by $\Delta\lambda_{\rm D}$), so a substantial part of the profile broadening and shape variability at low heights comes from Doppler shifts due to the LOS projection of the axial velocity. More inclined spicules yield larger projected speeds on average, and with azimuths sampled uniformly the LOS velocity distribution is approximately symmetric, producing comparable red- and blueshifted contributions.
An isolated spicule in our model does not produce an intrinsically double-peaked emission profile. The more complex profiles of the spicule forest, including double-peaked or M-shaped profiles in some realizations, arise from the LOS RT through many Doppler-shifted components with different velocities, source functions, and optical depths, including attenuation by foreground spicules.
This is consistent with the interpretation and multi-slab modeling of \citet{2020ApJ...888...42T}, who reproduced the low-height profile morphologies using symmetric LOS velocity distributions of order $\pm 25\,\mathrm{km\,s^{-1}}$. We further discuss this issue in Sect.~\ref{ssec:opdiam}.

\subsection{Effective spicule length}

The posterior expected effective length is $\approx 4\,\mathrm{Mm}$. We regard this as a conservative estimate. Substantially longer spicules remain compatible with our inference and are expected to appear in the data, with detectability generally favoring the more extended examples. Similar to Fig.~\ref{fig:mu}, in Fig.~\ref{fig:ell} we plot the mean Rayleigh density of $\ell$ together with several realizations drawn from the joint posterior of the hyperparameters.

As a check on geometric biases, we examined whether the effective length correlates with inclination by comparing $\mathbb{E}[\ell]$ with the mean cosine of the inclination, $\mathbb{E}(\mu)=\mathbb{E}[\cos\vartheta]$ (panel~H of Fig.~\ref{fig:posthyper}). We did not find any correlation between these quantities. We also studied the joint behavior of effective length and inclination by randomly selecting a posterior sample of the hyperparameters $(\sigma_\ell,\alpha_\mu,\beta_\mu)$ and drawing the associated spicule length and inclination cosine. We again find no correlation.

Finally, our current setup is not designed to model outliers (very long spicules). The Rayleigh family used for $\ell$ has a light tail and therefore down-weights rare, extreme lengths. Consequently, exceptionally long structures are not captured by the inferred distribution. Capturing that tail would require a heavier-tailed or mixture prior for $\ell$ and an analysis step tuned to rare, extended structures.

We note that the length $\ell$ in our forward model is measured from the spicule footpoints placed at the top of the chromosphere, at roughly $2\,\mathrm{Mm}$ above the visible limb, where the illuminating background is prescribed by the FAL-C spectrum. Therefore, lengths quoted here are referenced to this chromospheric boundary. When measured from the photosphere, the corresponding lengths would be larger by approximately this offset.

\subsection{Axial velocities}

\begin{figure}
\centering
\includegraphics[width=0.8\hsize]{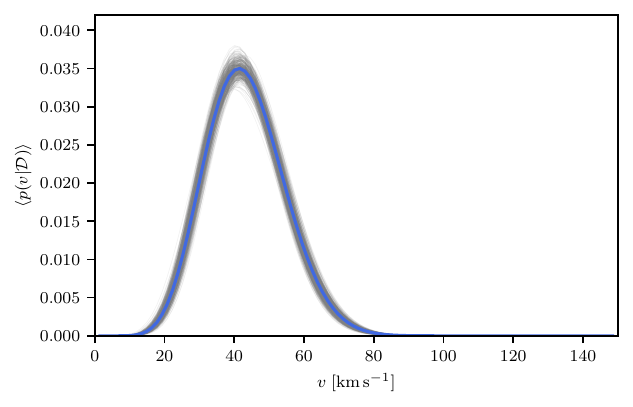}
\caption{
Distribution of the axial speed $v$ along the spicule axis.
}
\label{fig:v}
\end{figure}

\begin{figure}
\centering
\includegraphics[width=0.8\hsize]{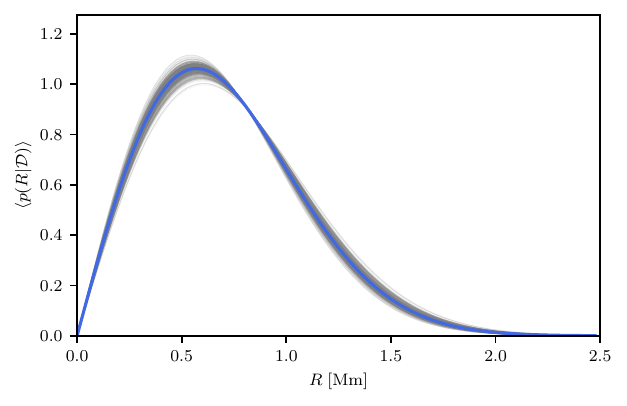}
\caption{
Distribution of the effective radius $R$.
}
\label{fig:fwhm}
\end{figure}

The longitudinal (axial) velocity was modeled with a beta prior on the normalized variable $v/v_{\max}\in[0,1]$. The joint posterior for $(\alpha_v,\beta_v)$ (panel~F of Fig.~\ref{fig:posthyper}) is well constrained and unimodal. The induced velocity distributions are shown in Fig.~\ref{fig:v}, where we plot the mean distribution together with several realizations drawn from the joint posterior of the hyperparameters. Very small axial velocities are excluded, the posterior expected value is $\mathbb{E}[v]\approx 50\,\mathrm{km\,s^{-1}}$, and there is nonnegligible posterior probability over roughly $20$–$80\,\mathrm{km\,s^{-1}}$.

This $\mathbb{E}[v]$ refers to motion along the spicule axis, whereas Doppler shifts measure the LOS component, i.e., a projection that is smaller on average, especially off the limb at the heights sampled here (Fig.~\ref{fig:obs}). Apparent axial speeds reported for type~II spicules span $\sim 50$–$150\,\mathrm{km\,s^{-1}}$ \citep[e.g.,][]{2007PASJ...59S.655D,2012ApJ...759...18P,2014ApJ...792L..15P}, while off-limb \ion{Mg}{ii} LOS velocities are typically $\lesssim 30\,\mathrm{km\,s^{-1}}$, with modeling that reproduces observed widths using random LOS velocities in the range $-25$ to $+25\,\mathrm{km\,s^{-1}}$ \citep{2020ApJ...888...42T}. In this context, a longitudinal mean near $50\,\mathrm{km\,s^{-1}}$ with most posterior support between $20$ and $80\,\mathrm{km\,s^{-1}}$ seems plausible and consistent with published ranges.

\subsection{Width of the line absorption profile
\label{ssec:width}}

We parameterized the absorption profile by the Doppler width $\Delta\lambda_{\rm D}$. The marginal posterior (panel~D of Fig.~\ref{fig:posthyper}) is sharply constrained with a maximum near $\Delta\lambda_{\rm D}\approx 0.037\,\text{\AA}$. If attributed purely to thermal broadening of the line, the corresponding kinetic temperature is $T\approx 2.3\times 10^{4}\,$K. Such a temperature is toward the upper end of chromospheric conditions typically associated with spicules \citep[e.g.,][]{2000SoPh..196...79S,2012SSRv..169..181T}. Around these temperatures, ionization of \ion{Mg}{ii} to \ion{Mg}{iii} becomes increasingly important, which would tend to reduce \ion{Mg}{ii} opacity \citep{2014A&A...564A.132H}. This suggests that part of the inferred $\Delta\lambda_{\rm D}$ likely arises from unresolved non-axial dynamics. Transverse swaying and torsional motions are commonly reported in spicules \citep{2012ApJ...752L..12D}. We therefore interpret $\Delta\lambda_{\rm D}$ as an effective width that combines thermal and unresolved dynamical broadening under our observing conditions.

We note that, in the present parameterization, $\Delta\lambda_{\rm D}$ is common to all spicules within a realization, while their inclinations and axial velocities are drawn independently from the corresponding population distributions. Correlations of the intrinsic line width with individual spicule properties are therefore not represented in the present model. At the population level, these quantities affect the spectra differently: $\Delta\lambda_{\rm D}$ controls the width of individual components, whereas inclination and axial velocity primarily control the distribution of their LOS Doppler shifts. A model allowing these properties to vary jointly would be a natural extension of the present framework.

\begin{figure}
\centering
\includegraphics[width=0.65\hsize]{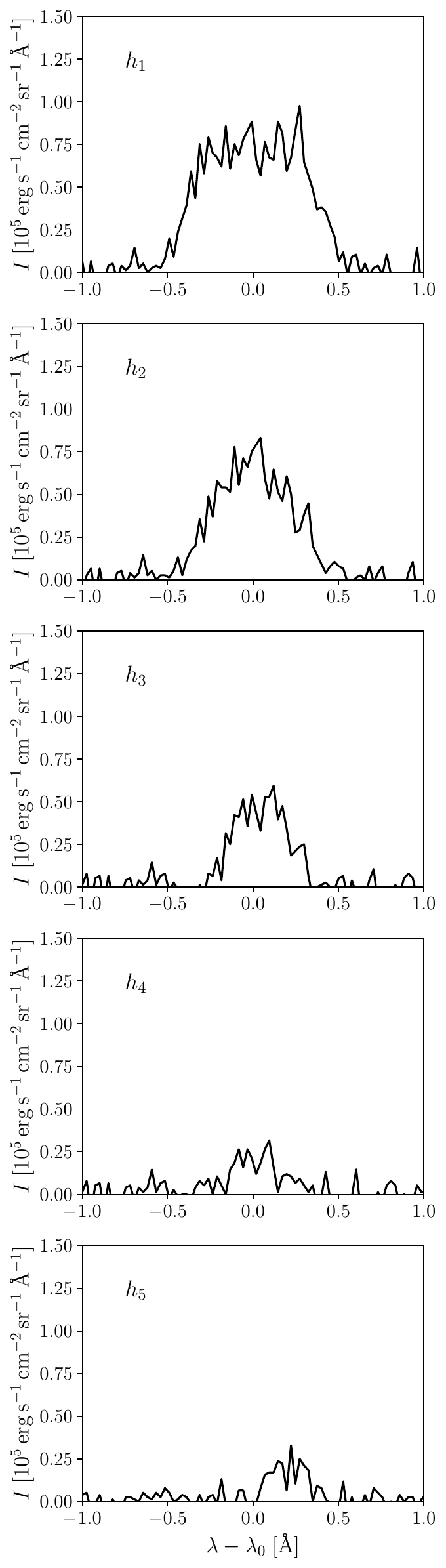}
\caption{
Synthetic profiles at five selected altitudes from Fig. \ref{fig:obs}, generated from a random realization of the spicular forest sampled from the MAP hyperparameter values of the hyperparameter distribution.
}
\label{fig:synth-prof}
\end{figure}

\begin{figure}
\centering
\includegraphics[width=0.65\hsize]{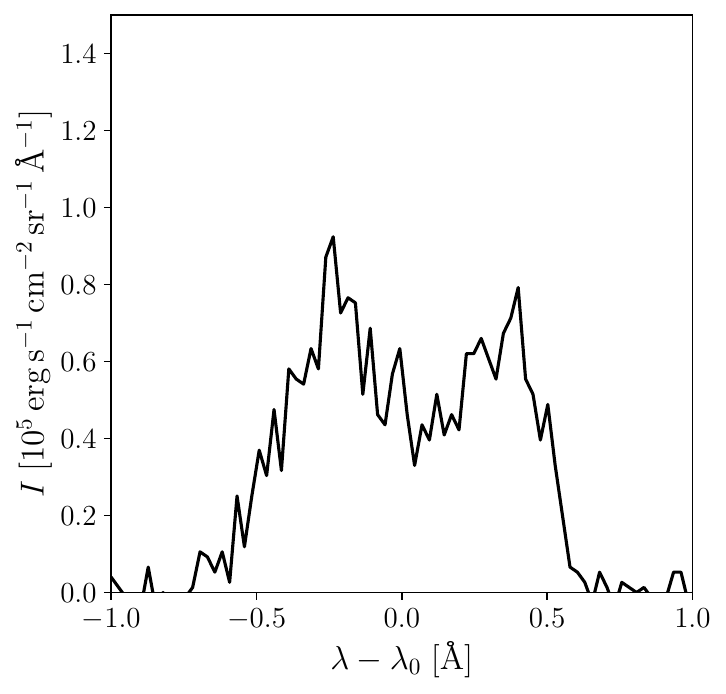}
\caption{
Example of a synthetic Mg II k line profile at an altitude of $2''$
above the top of the chromosphere, drawn from a random realization of the spicule forest at the MAP hyperparameter values. The profile exhibits the characteristic deep reversal (M shape) commonly observed in spicule spectra just above the chromosphere.
}
\label{fig:m-shape}
\end{figure}

\subsection{Opacity and effective diameter
\label{ssec:opdiam}}

We parameterized the line absorption with the opacity scale $\chi_{\rm L}$ and the effective diameter $D=2\sqrt{2\ln 2}R$ of a spicule. The marginal posterior of $\chi_{\rm L}$ (panel~C of Fig.~\ref{fig:posthyper}) is sharply constrained, with a maximum near $\chi_{\rm L}\approx 0.22\,\mathrm{Mm^{-1}}$. The posterior of the radius mode (panel~G of Fig.~\ref{fig:posthyper}) is likewise well constrained, peaking around $0.57\,\mathrm{Mm}$. Figure~\ref{fig:fwhm} shows the corresponding mean Rayleigh distribution together with several realizations drawn from the joint posterior of the hyperparameters. The corresponding effective diameter is then around $1.7\,{\rm Mm}$. We emphasize that $D$ is an effective transverse diameter in the RT model, not necessarily the geometric width of an individually resolved spicule. We address this issue in Sect.~\ref{sec:concl}.

Approximating the path integral in Eq.~(\ref{eq:tau0}) by an effective transverse width, the line-center optical thickness $\tau_0 \approx \chi_{\rm L} D$ becomes
\begin{equation}
\tau_0 \approx (0.22\, \mathrm{Mm^{-1}})\times(1.7 \,\mathrm{Mm}) \approx 0.37,
\end{equation}
i.e., a single spicule is moderately optically thin at line center in \mgk. This falls within the regime required by our radiative–transfer assumptions.

Figure \ref{fig:synth-prof} shows example synthetic spectra corresponding to the maximum a posteriori (MAP) solution, accounting for both instrumental spectral degradation and noise.

In the 1D NLTE slab calculations of \citet{2020ApJ...888...42T}, line-center optical thicknesses of order $\tau_0\sim 10$ arise for specific assumed thermodynamic states (e.g., $P=0.1\,\mathrm{dyn\,cm^{-2}}$, $T=10^4\,\mathrm{K}$, and $D=250\,\mathrm{km}$), while the parameter study shows that $\tau_0$ varies strongly with the adopted pressure, temperature, and geometrical thickness, so these inputs should be regarded as plausible but essentially ad hoc choices rather than empirically fixed spicule parameters. Moreover, optically thick, plane-parallel semi-infinite slabs are an idealized representation of spicules. A further difference concerns the number of spicules along the LOS. Given the large surface number densities, LOS near the limb will often intersect substantially more than ten spicules, a regime not captured by the typical $N\leq 10$ stacking used by \citet{2020ApJ...888...42T}. Consistent with these limitations, \citet{2020ApJ...888...42T} explicitly noted that their 1D configurations cannot reproduce the strongly reversed near-limb profiles. In our 3D population model, similarly M-shaped profiles, commonly observed just above the chromosphere, can appear in some realizations close to the chromosphere (see Fig.~\ref{fig:m-shape}), but not systematically, and their intensity still seems somewhat lower than observed. This indicates that the present model captures part of the relevant behavior but still lacks realism in the lowermost off-limb layers. We therefore view the $\tau_0$ estimates in \citet{2020ApJ...888...42T} and our effective $\tau_0\approx \chi_{\rm L}D$ as conditional results whose physical interpretation should be treated with caution.

\section{Discussion and conclusions
\label{sec:concl}}

Our inference framework treats spicules as a statistical ensemble rather than as individually fitted structures. This perspective yields joint posteriors for all hyperparameters, exposes parameter correlations, and provides uncertainty quantification. The spicule-population model used here is intentionally simple, but this is not a principal limitation of the inference strategy. The same likelihood-free approach can be applied to more physically informed models.

The approach also has limitations. As with any parametric modeling, the range of possible outcomes is constrained by the assumed families of probability distributions. Our results therefore describe the best-supported members of those families, not fully nonparametric shapes. In addition, we analyzed 250 line profiles per height. While adequate for a proof of concept, this sample may be insufficient to capture rare behavior or clear multimodality in the observed ensemble. Similarly, the PCA-based summaries emphasize the dominant variability across heights and are designed to be noise robust, but they may down-weight subtle profile-shape features with a diagnostic value for subpopulations.
Posterior-predictive comparisons of the observed and simulated profile distributions would provide a complementary test of the adequacy of the complete generative model and will be particularly valuable when applying the framework with more realistic forward models and alternative summary statistics.
The primary aim of the present work is therefore methodological, to demonstrate a population-level likelihood-free inference framework rather than to provide a definitive physical model of the spicule population.

A key interpretive point concerns the inclination distribution and the surface number density. The most frequently identified off-limb spicules appear predominantly vertical, although this partly reflects the projection and detection biases. More generally, spicules can be viewed as chromospheric structures that sometimes extend into the low corona. Since the chromosphere is not modeled explicitly here, some population parameters (most notably the inclination distribution and the surface number density) are best regarded as effective rather than literal. In particular, the inferred preference for stronger inclinations can be explained by the observational appearance: at the sampled heights, near-vertical spicules tend to stand out above the limb, whereas strongly inclined ones overlap with neighbors, remain within the low-lying forest, and are more likely to be missed or merged by an identification procedure.

One possible interpretation is that the transition region is strongly structured and that the upper chromosphere--corona interface is geometrically very complex. If so, the population that dominates the emission at lower heights may be more nearly horizontal than the population that remains distinguishable higher above the limb. In that case, a larger contribution from more inclined spicules in the lower layers would naturally enhance Doppler-shift projection effects. This is because even if each individual spicule has a comparatively narrow intrinsic absorption profile, the superposition of many azimuthally distributed, Doppler-shifted components can, in some realizations, produce M-shaped profiles similar to those commonly observed just above the chromosphere, although our model does not reproduce them systematically. This is consistent with the height-dependent \mgk\ morphology reported by \citet{2020ApJ...888...42T}. This may indicate that the inferred effective inclination and surface density encode a layered or corrugated chromosphere--corona interface rather than a single, height-independent geometry.

Within these limitations, the remaining inferred parameters are mutually consistent and physically plausible. The expected apparent length is $\mathbb{E}[\ell]\approx 4\,\mathrm{Mm}$, which we regard as a conservative estimate. Substantially longer spicules are not excluded by the inference and would generally be easier to detect observationally.

The mean apparent diameter is $D\approx 1.7\,\mathrm{Mm}$, which is larger than that commonly reported for individual spicules in high-resolution imaging and spectral-imaging studies. Modern measurements in chromospheric diagnostics such as \ion{Ca}{ii} and H$\alpha$ typically find widths of a few $10^2\ \mathrm{km}$, with the thinnest type-II spicules approaching the instrumental resolution limit \citep[$\lesssim 200\,\mathrm{km}$; e.g.,][]{2012ApJ...759...18P,2007PASJ...59S.655D}. Earlier ground-based compilations reported substantially larger diameters (up to 1--2.5 Mm), likely reflecting seeing and unresolved substructure \citep{1968SoPh....3..367B}. In our framework, $D$ is an effective scale that sets the LOS intersection length and optical thickness (and thus the emergent \mgk\ signal) at finite spatial resolution. It need not coincide with the geometric diameter of a single resolved spicule. However, it is still constrained by measurable radiative-transfer quantities, such as the typical LOS optical thickness. It should therefore follow the transverse extent of the \mgk-forming plasma, up to factors from unresolved groups of spicules and viewing geometry.

The Doppler width peaks at $\Delta\lambda_{\rm D}\approx 0.037\,\text{\AA}$, which, if interpreted as purely thermal for magnesium, corresponds to $T\approx 2.3\times 10^{4}\,\mathrm{K}$. However, a modest nonthermal contribution on the order of $2$--$3\,\mathrm{km\,s^{-1}}$ from unresolved transverse or torsional motions would bring this into the range of typical chromospheric conditions. Together with the diameter, the inferred opacity scale $\chi_{\rm L}\approx 0.22\,\mathrm{Mm}^{-1}$ implies a line-center optical thickness $0.37$ in \mgk\ on average, that is optically thin and consistent with our radiative-transfer regime. We stress that, given the simplicity of the model, these values are just order-of-magnitude estimates. 

The axial velocity distribution is well constrained and unimodal with $\mathbb{E}[v]\approx 50\,\mathrm{km\,s^{-1}}$ with most values between $20$ and $80\,\mathrm{km\,s^{-1}}$. Such speeds are plausible and naturally exceed the typically reported off-limb LOS Doppler speeds because the observations provide only the projected velocity components.

This work also illustrates why an ensemble-based inference is preferable to analyses of individual spicules in the forest. Tracing or fitting single structures requires the detection of spicules that are bright, isolated, and well separated. This, however, introduces selection effects and discards much of the signal carried by overlap and blending of individually undetectable spicules. By using all spectra at multiple heights and treating superposition and projection in the LOS explicitly, we obtain population parameters with credible intervals and parameter correlations that are directly tied to a radiative-transfer forward model. The resulting posteriors can be reused for posterior-predictive checks and as quantitative priors for subsequent forward modeling.

Several extensions are both desirable and physically motivated. On the diagnostic side, spectropolarimetry is needed to constrain magnetic fields at the chromosphere--corona interface, and multiline spectroscopy would help connect thermodynamic conditions across formation heights. On the modeling side, more realistic NLTE RT and additional thermodynamic degrees of freedom would help replace some of the present effective descriptors with more direct constraints on temperature, pressure, and ionization balance. On the inference side, alternative or complementary summary statistics could improve sensitivity to specific profile-shape differences, while more flexible population priors, such as mixture or semi-parametric families, would relax the present distributional assumptions. It would also be natural to extend the kinematic model, for example by explicitly representing separate upward and downward flow populations as well as non-axial motions, rather than capturing the former through a single unimodal axial-velocity distribution and the latter through the effective Doppler width.

Such an increase in model complexity would not, however, necessarily lead to more direct physical constraints. With the present \ion{Mg}{ii} k data, additional degrees of freedom may be weakly identifiable or mutually degenerate, so that some inferred quantities may remain effective. More complex models will therefore be most useful when combined with additional observational constraints, such as multiline spectroscopy and spectropolarimetry.

More generally, the overall inference framework could in principle be adapted to other ubiquitously occurring and overlapping small-scale solar features, such as fine structure in prominences and filaments. However, the corresponding population models and prior distributions would likely differ substantially from those adopted here for spicules.

In practice, the increased computational cost of such extensions may be significant but not prohibitive. The population inference needs to be carried out only once per representative phase of the solar cycle. In future work, it would be preferable to target the average quiet Sun rather than a coronal hole, since the quiet Sun is closer to a statistically stationary regime and should yield more stable population posteriors.

In summary, we do not make definitive claims about spicule parameters. Instead, we propose a generalizable statistical route to retrieve population properties. Different formulations of the forward model, richer diagnostics, and more flexible priors may lead to more decisive constraints. In this context, forthcoming high-resolution spectropolarimetric data from the balloon-borne Sunrise~III mission \citep{2025SoPh..300...75K} should provide the type of multiband spectropolarimetric constraints needed to test and refine our framework.

\begin{acknowledgements}
J.\v{S}, S.G., and P.H. acknowledge the financial support from the grant 25-18282S of the Czech Science Foundation (GA\v CR) and the project \mbox{RVO:67985815} of the Astronomical Institute of the Czech Academy of Sciences.
PH was supported by the program `Excellence Initiative -- Research University' for the years 2020--2026 for the University of Wroc\l aw, project no. BPIDUB.4610.96.2021.KG.
This work was supported by JSPS KAKENHI Grant Numbers JP23KJ2151 (PI: A.T.) and JP21K13971 (PI: A.T.).
\end{acknowledgements}

\bibliographystyle{aa}
\bibliography{ms.bib}

\end{document}